\documentclass[aps,amsmath,twocolumn,amssymb,floatfixng,showpacs,superscriptaddress,footinbib]{revtex4-1}
\pdfoutput=1
\usepackage[dvips]{graphics}
\usepackage{hyperref}
\usepackage{bm}
\usepackage{epsfig}
\usepackage{enumerate}
\usepackage{color}
\usepackage{graphicx}
\definecolor{dred}{rgb}{0.6,0,0}
\hypersetup{
    pdfnewwindow=true,      
    colorlinks=true,       
    linkcolor=magenta,          
    citecolor=blue,         
    filecolor=blue,      
    urlcolor=blue        
}
\newcommand{\bea}{\begin{eqnarray}}
\newcommand{\eea}{\end{eqnarray}}
\newcommand{\beq}{\begin{equation}}  
\newcommand{\eeq}{\end{equation}}
\newcommand{\non}{\nonumber}

\newcommand{\etal}{{\it{et al.,~}}}

\newcommand\ie{i.e.,\,}
\begin{document}
\title{Thermoelectricity carried by proximity-induced odd-frequency pairing in ferromagnet/superconductor junctions}
\author{Paramita Dutta}
\email{paramita.dutta@physics.uu.se}
\affiliation{Department of Physics and Astronomy, Uppsala University, Box 516, S-751 20 Uppsala, Sweden}
\author{Kau\^{e} R. Alves} 
\affiliation{Department of Physics and Astronomy, Uppsala University, Box 516, S-751 20 Uppsala, Sweden}
\affiliation{Department of Mathematical Physics, Institute of Physics, University of S\~{a}o Paulo, CEP 05314-970, 
S\~{a}o Paulo, Brazil}
\author{Annica M. Black-Schaffer}
\affiliation{Department of Physics and Astronomy, Uppsala University, Box 516, S-751 20 Uppsala, Sweden}

\begin{abstract}
We explore the role of proximity-induced odd-frequency pairing in the thermoelectricity of a ferromagnet when coupled to a conventional $s$-wave spin-singlet superconductor through a spin-active interface. By varying both the polarization and its direction in the ferromagnet and the interfacial spin-orbit interaction strength, we analyze the behavior of all proximity-induced pair amplitudes in the ferromagnet and their contributions to the thermoelectric coefficients. Based on our results for the Seebeck coefficient, we predict that odd-frequency spin-triplet Cooper pairs are much more efficient than the conventional spin-singlet even-frequency pairs in enhancing thermoelectricity of the junction, and especially mixed-spin triplet pairing is favorable.  Our results on the thermoelectric figure of merit show that ferromagnet/superconductor junctions are very good thermoelectric systems when superconductivity is dominated by odd-frequency pairing.
\end{abstract}

\maketitle

\section{Introduction}
The phenomenon of odd-frequency (odd-$\omega$) superconductivity occurs when the superconducting pair expectation value is odd under the exchange of time, or equivalently frequency, of the two electrons in the Cooper pair\,\cite{kirkpatrick1991tr,BalatskyNewclass,Schrieffer1994,linderRMP}. Following the first prediction by Berezinskii\,\cite{Berezinskii} in the context of $^3$He, the concept of odd-$\omega$ pairing was subsequently introduced for superconductivity \,\cite{kirkpatrick1991tr,BalatskyNewclass,Schrieffer1994}. This unusual pairing obeys some exotic symmetries, such as $s$-wave spin-triplet and $p$-wave spin-singlet symmetry\,\cite{linderRMP}. It has mostly been found in hybrid structures like ferromagnet (FM)/superconductor (SC)\,\cite{linder2015superconducting,linder2015strong,Rmp,di2015a,tanakaFS,Matsumoto2013}, normal metal/SC\,\cite{TanakaNSjunction,qtLinder,CayaooddwRashba,cayao2020odd}, and multiband systems with inter-band hybridization\,\cite{Annicamulti,komendova2017odd,dutta2019finite,triola2020role}. 

Several attempts have been made to experimentally identify odd-$\omega$ pairing. Large efforts have been concentrated towards the indirect detection via proximity effect and Josephson current\,\cite{keizer2006,Khaire2010,robinson2010}. Later, other manifestations of odd-$\omega$ superconductivity have also been reported using scanning tunneling measurement\,\cite{di2015a} and the paramagnetic Meissner effect\,\cite{di2015b} following several theoretical predictions\,\cite{BalatskyMeissner,TanakaJosephson,TanakaMeissner2}. There also exist other proposals  based on Josephson current\,\cite{parhizgar,Duttawnls} and Kerr effect\,\cite{komendova2017odd} for the detection of odd-$\omega$ pairing in unconventional superconductors. In all the above-mentioned work, the electron transport properties are used to identify and understand the role of odd-$\omega$ pairing. It is also interesting to look at the thermal transport phenomena in the SC hybrid junctions, especially FM/SC interfaces, which are both excellent hosts of odd-$\omega$ pairing\,\cite{linder2015superconducting,linder2015strong,Rmp,di2015a,tanakaFS,Matsumoto2013} and also identified as well-behaved thermoelectric junctions\,\cite{ozaeta2014,machon,kolenda2016}.

In general, SCs are not good thermoelectric materials in comparison to normal metals as the supercurrent easily interferes with thermal current making it hard to isolate and thus utilize\,\cite{galperin}. Even in the case of successful isolation, thermoelectricity in SCs is weak because of the particle-hole symmetric energy spectrum of conventional SCs in the linear regime\,\cite{machon,Giazotto}. However, thermoelectricity can be enhanced by breaking the particle-hole symmetry for each spin separately\,\cite{ozaeta2014}, due to the resulting asymmetry in the energies of the two spin bands\,\cite{Tedrow}. This can be achieved by locally applying a spin-splitting field forming a FM region and then proximity couple this FM to a SC constructing a hybrid structure\,\cite{ozaeta2014}. 

Recently, the idea of implementing spin-splitting through FM/SC structures has been shown to yield considerable enhancement of the thermoelectricity.\,\cite{chandrasekhar2009,kalenkov2012theory,ozaeta2014,machon,kolenda2016,kolendaprl2016,PDFM,kamra,linder2016}. Such enhancement of the thermoelectricity is always useful due to the prospects of application\,\cite{Thermometers,Detector,shakouri,islam}. 
Experimental observation of thermoelectricity in FM/SC structures is a big step forward in this direction. The theoretically predicted very large thermoelectric effect in FM/SC junctions was experimentally observed in 2016 with excellent agreement to theory  using a high magnetic field \cite{kolendaprl2016}. The requirement of large magnetic field was subsequently eliminated by using a ferromagnetic insulator a year later \cite{kolendaprb2017}. Very recently, nonlinear thermoelectric effects have also been experimentally observed in FM/SC structures \cite{beckmannprb2019,kolenda2016}.

However, in works on thermoelectricity in FM/SC interfaces, mainly the role of the conventional $s$-wave spin-singlet even-frequency (even-$\omega$) pairing has been discussed, although odd-$\omega$ pairing is also inherently present in these systems, often even dominating over even-$\omega$ pair amplitudes. Thus, question arises: what is the role of odd-$\omega$ pair amplitude in thermoelectricity in FM/SC structures? The question is highly relevant as Hwang \etal has proposed thermoelectricity as a way to detect the odd-$\omega$ superconductivity in quantum dot systems\,\cite{sothman}. More specifically, they have revealed that in a FM/Quantum-dot/SC system one of the thermoelectric coefficients, the so-called thermal coefficient which measures the charge current induced by a temperature gradient, is zero for even-$\omega$ pairing but finite for odd-$\omega$ pairing and thus, thermoelectricity can be used as a probe for the odd-$\omega$ pairing in the quantum dot. Additionally, recently Keidel \etal have proposed a way to generate equal-spin triplet Cooper pairs at the helical edge states of a quantum spin Hall insulator and used that to drive a supercurrent from a temperature gradient in a SC/FM insulator/SC structure along the edge\,\cite{keidel2019ondemand}. Despite these interesting results, it is not yet established whether odd-$\omega$ pairing is a good carrier of thermoelectricity in generic FM/SC structures. More specifically, is thermoelectricity in FM/SC junctions enhanced in the presence of odd-$\omega$ pairing compared to the scenario when there is only even-$\omega$ pairing in the system? Or, in other words, how efficient are odd-$\omega$ Cooper pairs as carriers of the thermoelectric current to make the FM/SC junction an efficient thermoelectric system? Moreover, the existence of a finite density of states within the SC gap due to odd-$\omega$ superconductivity\,\cite{asano2008,foglstrom} is also interesting, as that might influence the subgap contributions to the thermal current, which is absent in the case of even-$\omega$ pairing.

Motivated by this, we study FM/SC junctions and explore the behavior of an experimentally measurable quantity, the Seebeck coefficient or thermopower\,\cite{blundell2009concepts}, which provides the ability of heat transfer through the junction. The interfacial region is considered to host a spin-active region from Rashba spin-orbit interaction (RSOI)\,\cite{rashba,rashba2}, which  also produces odd-$\omega$ spin-triplet pairing in an efficient way\,\cite{eschirg}. To explain the role of odd-$\omega$ pairing, we first analyze the proximity-induced pair amplitude in the FM/SC junction. Then, we show that the magnitude of the Seebeck coefficient is enhanced when the polarization of the FM is equal or close to one and the RSOI strength is finite, which is also when the proximity-induced odd-$\omega$ spin-tripet pairing is dominating over even-$\omega$ states. In particular, we show that the subgap contribution to the Seebeck coefficient is both large, and, most importantly, caused by odd-$\omega$ pairing. We also show that mixed-triplet spin pairing seems to be more efficient than equal-spin triplet pairing in enhancing the thermoelectric effect. Enhancement of the Seebeck coefficient indicates the possibility of getting good thermopower based on odd-$\omega$ superconductivity. This prediction is supported by our result of the thermoelectric figure of merit $zT$, characterizing the efficiency of the junction. For any thermoelectric material, it is hard to achieve a value of $zT$ more than 1, the value which is well-known as an indicator for an efficient thermoelectric material. We show that $zT$ comes out to be as high as 5 in the parameter regime where the ratio of odd- to even-$\omega$ pair amplitude is large. Overall, this leads to the conclusion that odd-$\omega$ pairing yields significant contributions in enhancing the thermoelectricity of FM/SC junctions. 

We organize the rest of the article as follows. In Sec.\,\ref{mt} we present our model for FM/SC junctions. The theory and the analysis of the pair amplitude are discussed in Sec.\,\ref{Pamp}. Our results for the thermoelectric coefficients are then presented in Sec.\,\ref{thermo}, including the necessary theoretical background. Finally, we summarize our results in Sec.\,\ref{conclu}.

\section{Model}\label{mt}
\begin{figure}[htb]
\begin{center}
\includegraphics[scale=1.1]{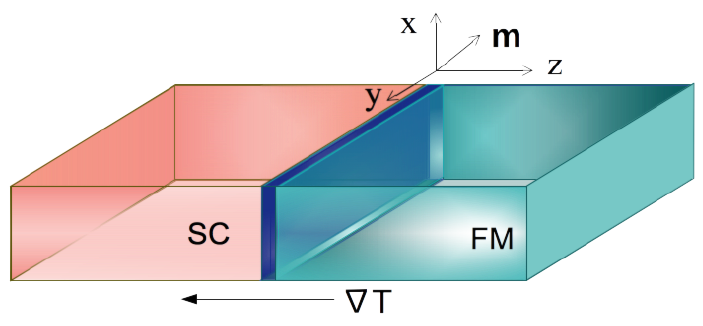}
\caption{Schematic picture of the FM/SC junction with magnetization vector $\mathbf{m}$ and interfacial spin-active region marked by thin blue layer. An infinitesimal temperature gradient $\nabla T$ is applied across the junction.}
\label{model}
\end{center}
\end{figure}

We consider a FM attached to a conventional $s$-wave spin-singlet SC with a very thin spin-active interface in between, as shown schematically in Fig.\,\ref{model}. Initially, the system is at an equilibrium temperature $T$ and then a temperature gradient $\nabla T$ is applied across the junction. We describe each part of the junction with the Bogoliubov-de Gennes (BdG) equation\,\cite{de1999superconductivity},
\beq
H_{\xi}({\bm k}) \Psi({\bm k}) = E \Psi({\bm k})
\eeq
with
\beq
H_{\xi}({\bm k})=
\begin{pmatrix}
H^0_{\xi}({\bm k}) & \Delta_{\xi}\, \sigma_y \\
\Delta_{\xi}^\dagger\, \sigma_y & -H_{\xi}^{0\,*}(-{\bm k})
\end{pmatrix},
\label{Hmat}
\eeq
where $H^0_{\xi}({\bm k})$ represents the normal part of  either the FM Hamiltonian $H^0_{\text{FM}}({\bm k})$ for $z>0$ or 
the SC Hamiltonian, $H^0_{\text{SC}}({\bm k})$ for $z<0$. $\Delta_{\xi}$ is the gap parameter which is zero for in the FM region and $\Delta_\text{SC}$ for the SC region given by 
\beq
\Delta_{\text{SC}}=\Delta_0 \tanh ( 1.74 \sqrt{T_c/T-1}),
\eeq
where $T_c$ is the critical temperature of the SC. This is notably the only property where temperature effects enter explicitly in the Hamiltonian.

The single-particle Hamiltonian for the FM including the spin-active interface is taken as
\bea
H^0_{\text{FM}}(\bm{k})&=&\hbar^2\bm{k}^2/2-(h_{\text{FM}}/2)~\mathbf{m}\cdot \bm{\sigma}-\mu_{\text{FM}}\nonumber \\
&&~~~~~~~~~~~~~~~~~~~~~~~~~~~+\mathbf{W_{\text{RSOI}}}\cdot {\bm \sigma}\, \delta(z)
\label{ham_fm}
\eea
where the first term describes the kinetic energy, the second term introduces the magnetism, the third term is the chemical potential of the FM, and the fourth term represents the interfacial spin-active region. The magnetism is expressed through an exchange field with magnitude $h_{\text{FM}}$ and direction $\mathbf{m}=\{\sin{\theta_{\text{F}}}\cos{\phi}_{\text{F}},\sin{\theta_{\text{F}}}\sin{\phi_{\text{F}}},\cos{\theta_{\text{F}}}\}$ defined by the polar angle $\theta_{\text{F}}$ and azimuthal angle $\phi_{\text{F}}$\,\cite{stoner1939collective}. Here, ${\bm \sigma}$ represents the Pauli matrices for the spin degree of freedom. The Rashba field ${\mathbf W}_{\text{RSOI}}$ of the interfacial spin-active region is chosen as $\lambda_{\text{RSOI}}[k_y,-k_x,0]$ assuming a growth direction of the heterostructure along [001] crystallographic axis\,\cite{matos2009angular}. Here, the RSOI field strength is denoted by the parameter $\lambda_{\text{\text{RSOI}}}$. 
The normal part of the SC Hamiltonian is taken as
\bea
H^0_{\text{SC}}(\bm{k})&=&\hbar^2\bm{k}^2/2-\mu_{\text{SC}}.
\label{ham_sc}
\eea
with the first term representing the kinetic energy followed by the chemical potential of the SC. 

For the illustration of our results, we define some dimensionless parameters and use those throughout the work. The RSOI strength is scaled as $\Lambda_{\text{RSOI}}$$=$$\frac{2\lambda_{\text{RSOI}}}{\hbar^2}$ and the spin polarization in the FM is redefined as $P$$=$$\frac{h_{\text{FM}}}{2 \Delta_0}$. Further, we set $\hbar=1$, $\Delta_0=0.1$ and $T/T_c=0.5$, although our results are valid for any $T/T_c<1$ including the experimentally accessible temperatures. With these parameter values, we can write the scattering matrix equations using Eq.\,(\ref{Hmat}) and find the thermoelectric coefficients as described in Sec.\,\ref{thermo}. 

In order to understand in depth the behavior of the pair amplitude, we consider the real space (in one direction) Hamiltonian of the whole FM/SC structure with a tunnel coupling in between the FM and SC regions. For this, we discretize the BdG Hamiltonian (using Eqs.\,(\ref{Hmat}-\ref{ham_sc})) by taking the inverse Fourier transformation along $z$-axis, but keep the periodicity along $x$- and $y$-axis. After coupling the SC and FM by a tunneling term, the discretized BdG Hamiltonian for the whole FM/SC junction takes the form
\bea
H_{\text{FS}}(\bm{k_{||})}&=&H_{\text{FM}}+H_{\text{SC}}+H_{\text {FM-SC}} \non \\
&=&\sum_{n\in \text{FM},\sigma}[\bm{c}_{n,\sigma}^{\dagger}(2-\cos k_x-\cos k_y+P\,\bm{m}.\bm{\sigma} \non\\
&&+\mu_{\text{FM}}+\Lambda_{\text{RSOI}}\, (\sigma_x\sin k_y- \sigma_y\sin k_x)\delta_{i,0}) \bm{c}_{n,\sigma}\non\\
&&+t_{\text{FM}} (\bm{c}_{n,\sigma}^{\dagger}\bm{c}_{n+1,\sigma}+\text{H.c.})] \non \\
&&+\sum_{l\in \text{SC},\sigma,\sigma^{\dagger}}[\bm{b}_{l,\sigma}^{\dagger}(2-\cos k_x-\cos k_y+\mu_{\text{SC}}) \bm{b}_{l,\sigma}\non \\
&&+(t_{\text{SC}}\bm{b}_{l,\sigma}^{\dagger}\bm{b}_{l+1,\sigma}+\Delta_{\text{SC}} \bm{b}^{\dagger}_{l,\sigma}\bm{b}_{l,\sigma^{\prime}}^{\dagger} +\text{H.c.})]\non\\
&&+ \sum_{\langle n\in\text{FM},l\in\text{SC}\rangle,\sigma}t_{\text{FM-SC}}(\bm{c}_{n,\sigma}^{\dagger}\bm{b}_{l,\sigma}+\text{H.c.}),\eea
where $n$, $l$ index the different layers along the $z$-axis. $\bm{c}^{\dagger}_{n,\sigma}$ ($\bm{b}^{\dagger}_{l,\sigma}$) and $\bm{c}_{n,\sigma}$ ($\bm{b}_{l,\sigma}$) are the creation and annihilation operators for the $n$ ($l$)-th layer within the FM (SC). The $\langle..\rangle$ term denotes the nearest neighbor term. We also skip the wave vector notation $\bm{k}_{||}$ ($k_x, k_y$) (\ie parallel to the interface) from all operators for the sake of compactness. Moreover, $t_{\text{FM}}$ ($t_{\text{SC}}$) is the nearest neighbor hopping between the adjacent layers of the FM (SC) and the coupling between FM and SC is denoted by $t_{\text{FM-SC}}$. We note that this FM-SC coupling is the only additional parameter needed in the lattice formalism compared to the continuum formalism in Eqs.\,(\ref{Hmat}-\ref{ham_sc}). In the continuum model this interface coupling is incorporated into the boundary conditions and thus changing this interface hopping parameter only quantitively affects our results (as would changing the boundary conditions to a slightly less transparent junction). Most importantly, none of our results is qualitatively sensitive to its value. The total number of layers in FM and SC are denoted by $N_{\text{FM}}$ and $N_{\text{SC}}$, whereas $N_{\text I}$ is the number of interface layer(s) at (around) $z=0$ with RSOI. We show all the results for the pair amplitudes for $N_{\text{FM}}=N_{\text{SC}}=50$ and $N_{\text{I}}=2$. To keep the model simple, we consider $t_{\text{FM}}=t_{\text{SC}}=1$ and $t_{\text{FM-SC}}=0.5$. For the chemical potential, we take $\mu_{\text{FM}}=0$ and $\mu_{\text{SC}}=2$, though the main results are not qualitatively sensitive to the numerical values of the parameters. Changing $t$ values will scale the pair amplitude keeping the qualitative behavior the same. Also, our system size for the FM/SC junction is sufficiently large as the SC coherence length is $10 a$ with $a$ being the lattice constant.

\section{Pair Amplitude} \label{Pamp}
In order to understand the proximity-induced superconductivity in the FM region, we analyze all pair amplitudes, including odd-$\omega$ pairing, in the FM region by calculating the anomalous Green's function for the whole FM/SC junction as discussed in the following subsections.
\subsection{Theoretical background} 
The Cooper pair amplitude in superconductors can be found from the anomalous Green's function defined as the time-ordered expectation value of the field operators for the fermions with spins $\sigma$ and $\sigma^{\prime}$, 
\bea
F_{\sigma,\sigma^{\prime}}(r,t)&=-&\langle\mathcal{T}_t \bm{\Psi}_\sigma(r,t) \bm{\Psi}_{\sigma^{\prime}}(r,0)\rangle \non \\
&=&-\Theta(t)\langle \bm{\Psi}_\sigma(r,t) \bm{\Psi}_{\sigma^{\prime}}(r,0)\rangle\non \\
&&+\Theta(-t)\langle \bm{\Psi}_{\sigma^{\prime}}(r,0) \bm{\Psi}_{\sigma}(r,t)\rangle,
\label{pamp}
\eea 
where $\mathcal{T}_t$ is the time-ordering operator and $\Theta(t)$ is the heaviside step function. 
The time-dependent field operator $\bm{\Psi}_{\sigma}(r,t)$ can be found from the Heisenberg picture as
\bea
\bm{\Psi}_{\sigma}(r,t)=e^{i H_{\text{FS}}(\bm{k}_{||})t}~\bm{\Psi}_{\sigma}(r,0)~e^{-i H_{\text{FS}}(\bm{k}_{||}) t},
\label{heisenberg}
\eea
where the time-independent form $\bm{\Psi}_{\sigma}(r,0)$ may be either $\bm{c}_{r,\sigma}$ or $\bm{b}_{r,\sigma}$. 

The Fourier transform of the anomalous Green's function $F_{\sigma,\sigma^{\prime}}(r,t)$ provides the pair amplitude in frequency space $\mathcal{F}_{\sigma,\sigma^{\prime}}(r,\omega)$, 
\begin{figure*}[htb]
\begin{center}
\includegraphics[scale=0.68]{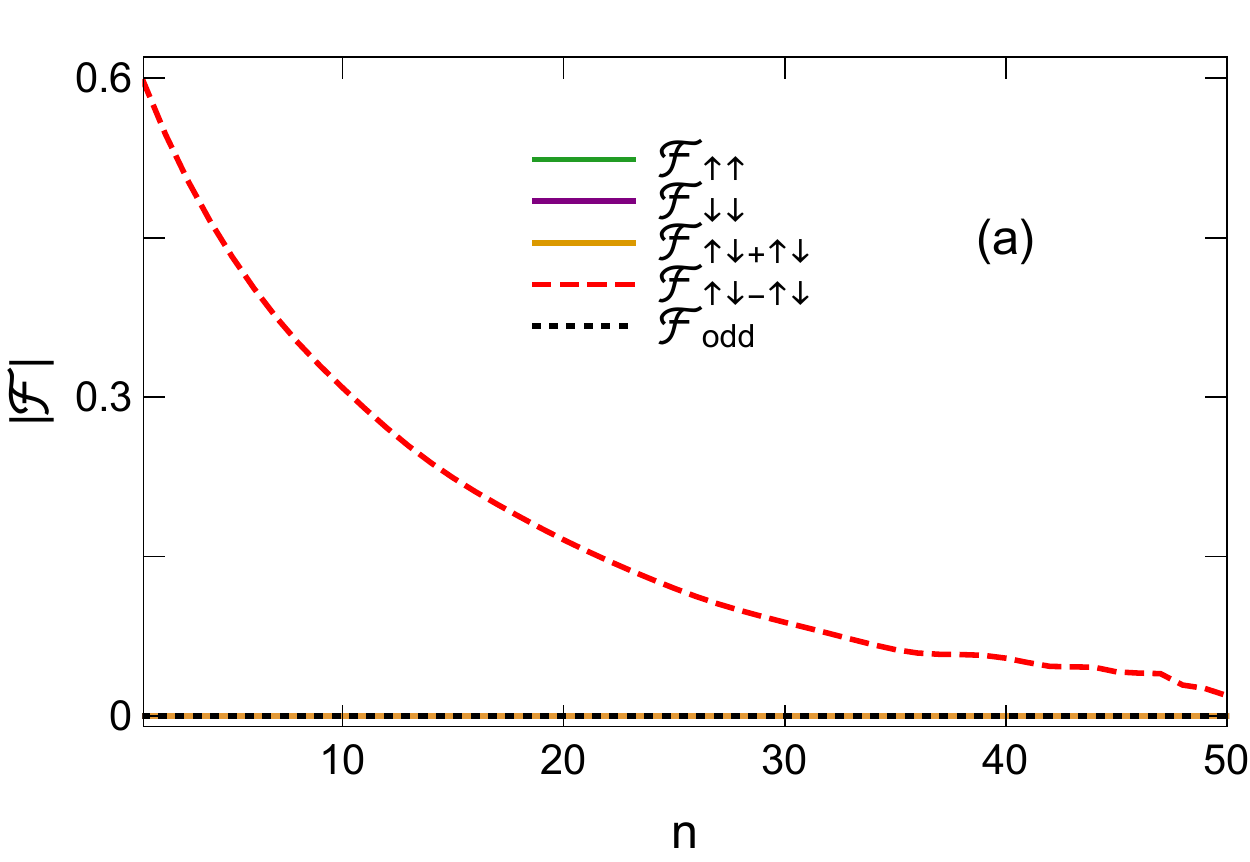}
\includegraphics[scale=0.68]{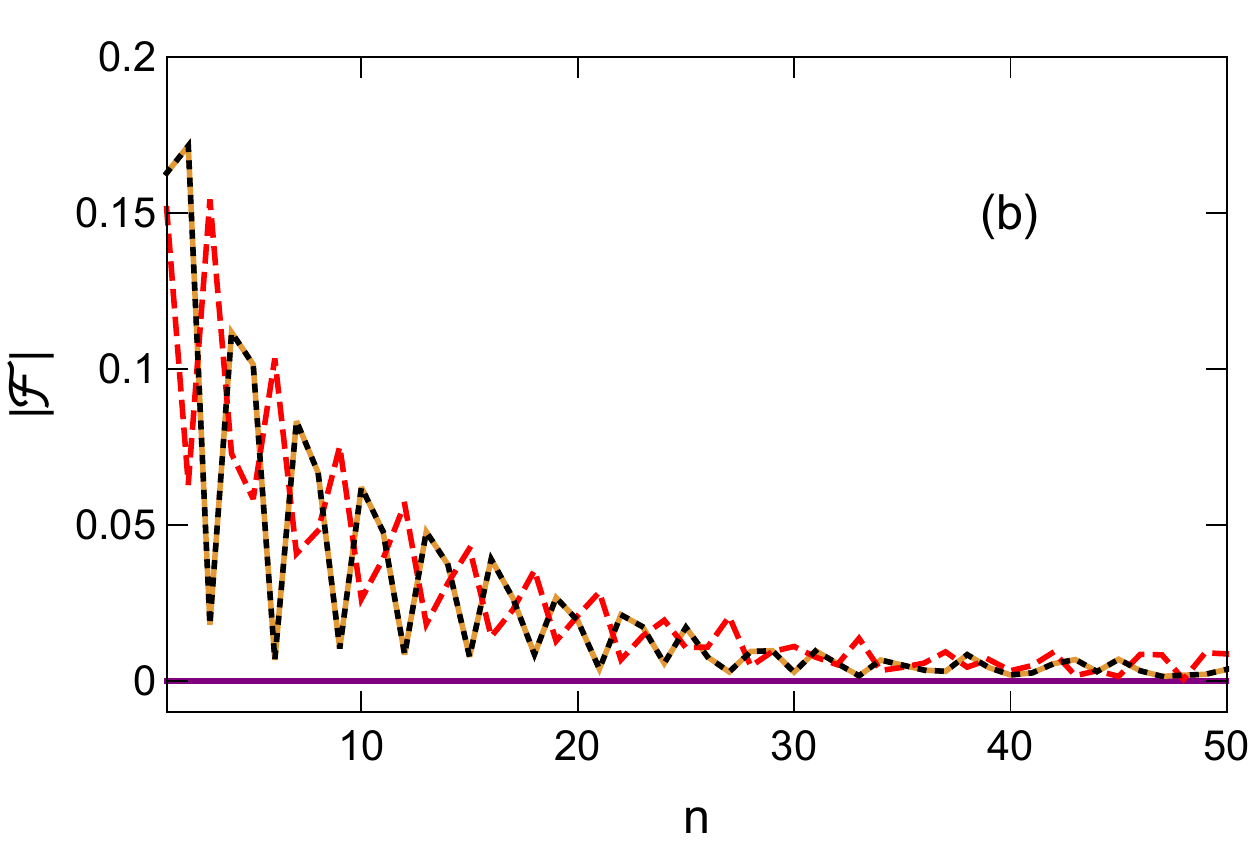}\\
\includegraphics[scale=0.68]{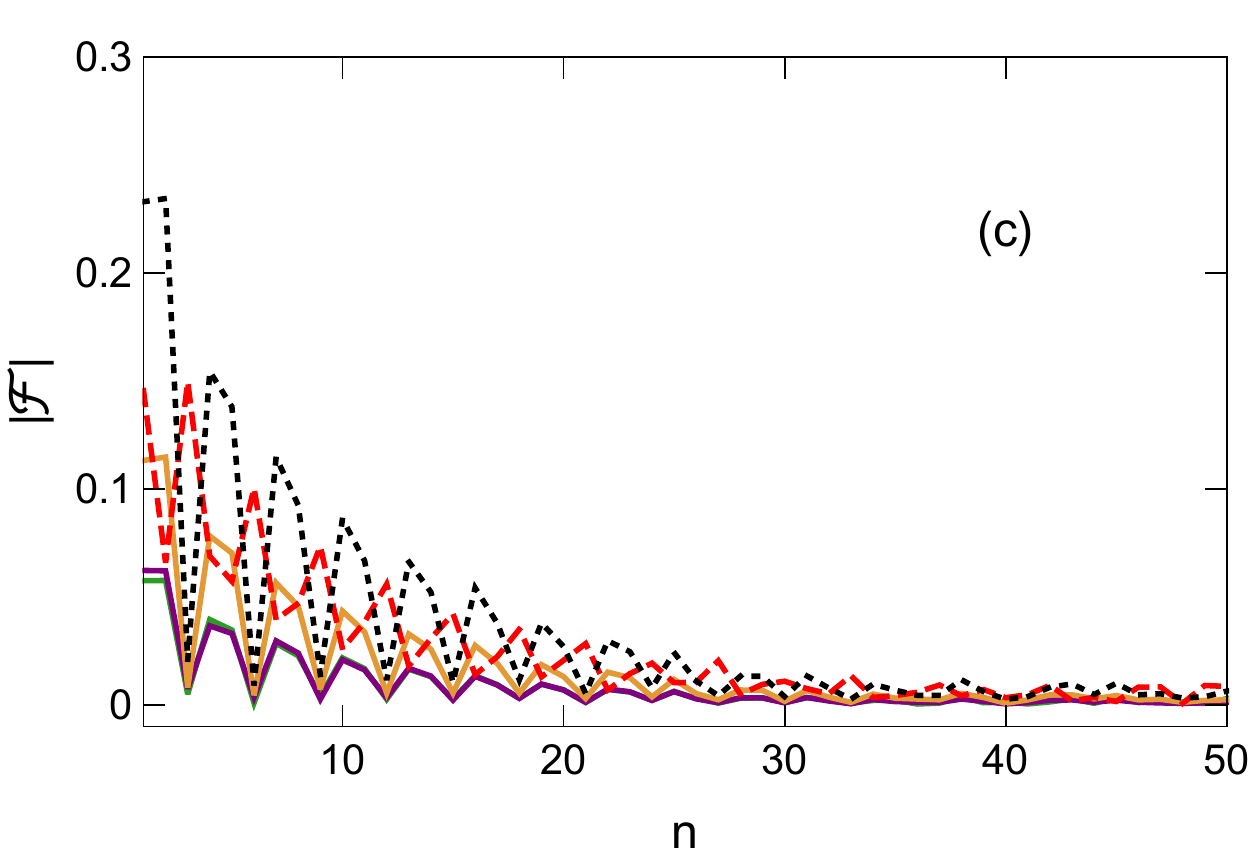}
\includegraphics[scale=0.68]{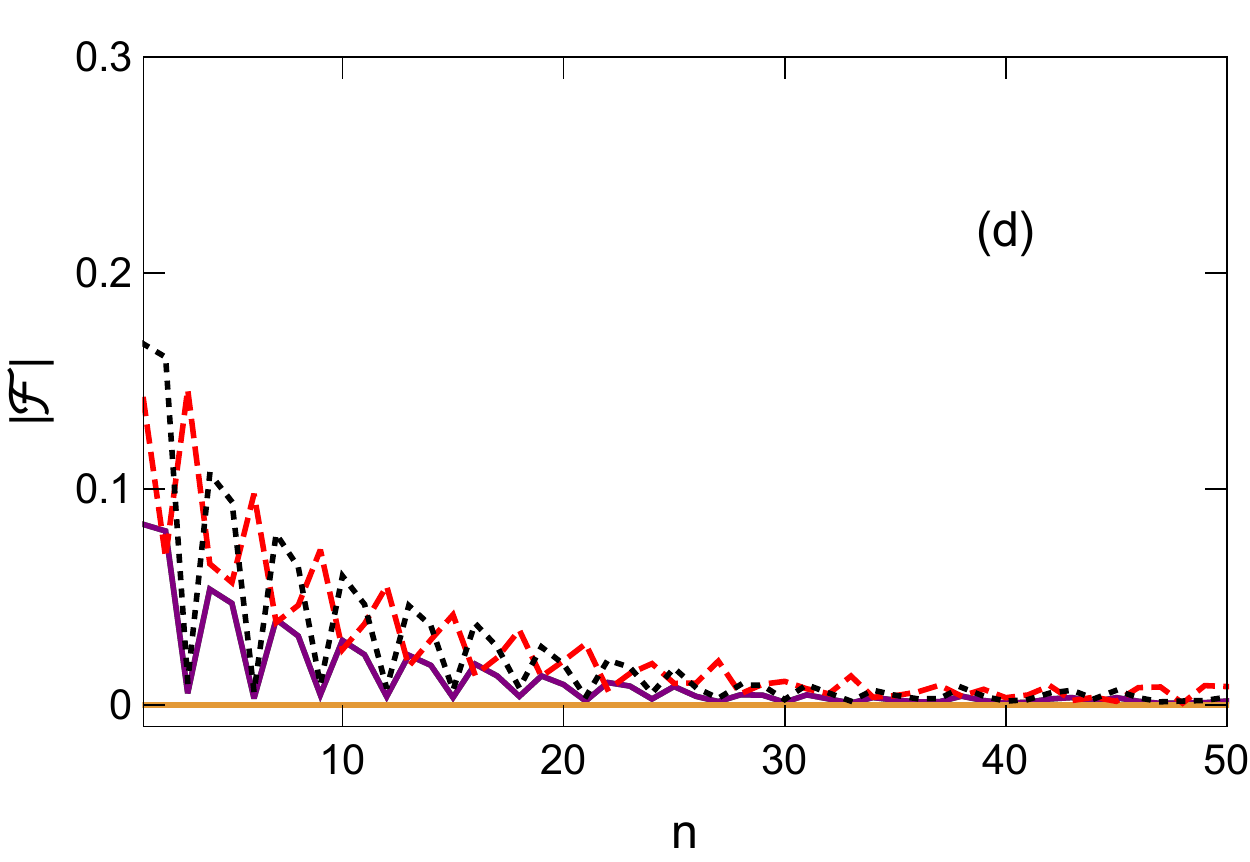}
\caption{Pair amplitude $|\mathcal{F}|$ as a function of site index in the FM in (a) the  absence of RSOI ($\Lambda_{\text{RSOI}}=0$) and (b)-(d) presence of RSOI ($\Lambda_{\text{RSOI}}=0.3$). The polarization is set as (a) $P=0$, $\theta_{\text F}=0$, (b) $P=1$, $\theta_{\text F}=0$, (c) $P=1$, $\theta_{\text F}=\pi/4$, and (d) $P=1$, $\theta_{\text F}=\pi/2$, keeping $\phi_{\text F}=0$ for all panels. All four spin configurations are mentioned in the legend, and in addition $\mathcal{F}_{\text{odd}}$ is the sum of all the spin-triplet odd-$\omega$ pair amplitudes.}
\label{figPA}
\end{center}
\end{figure*}
which we calculate using Green's function of the junction as follows\,\cite{BalatskyMeissner}: 
We start by defining the retarded Green's function of the whole FM/SC junction as
\bea
G^R(\omega,\bm{k}_{||})=[(\omega+i \eta) I-H_{\text {FS}}({\bm{k}_{||}})]^{-1}.
\label{green}
\eea
$\eta$ is an infinitesimal quantity and $I$ is the identity matrix.
The anomalous part of the Green's function is found from the block-matrix form of the Green's function in the Nambu basis as
\beq
G^R(\omega,\bm{k}_{||})=
\begin{pmatrix}
\mathcal{G}(\omega,\bm{k}_{||}) & \mathcal{F}(\omega,\bm{k}_{||}) \\
\bar{\mathcal{F}}(\omega,\bm{k}_{||}) & \bar{\mathcal{G}}(\omega,\bm{k}_{||})\end{pmatrix},
\label{Gmat}
\eeq
where each component of the block-matrix $G^R(\omega,\bm{k}_{||})$ is a $2N \times 2N$-dimensional matrix, $N$ being the total number of the layers ($N=N_{\text{FM}}+N_{\text{SC}}+N_{\text I}$) in the whole FM/SC junction along the $z$-axis. 
Furthermore, the anomalous part of the Green's function can be expressed as
\bea
\mathcal{F}(\omega)
=\sum\limits_{\bm{k}_{||}}\mathcal{F}(\omega,\bm{k}_{||}),
\label{Fw}
\eea
where we take the summation over the planar wave vector $\bm{k}_{||}$ (which is a good quantum number) within the first Brillioun zone, to account for the periodicity along the $x$- and $y$-axis. The matrix form of $\mathcal{F}$ looks like
\bea
\mathcal{F}(\omega)=
\begin{pmatrix}
\mathcal{F}_{\uparrow\uparrow}(\omega) & \mathcal{F}_{\uparrow\downarrow}  (\omega)\\
\mathcal{F}_{\downarrow\uparrow}(\omega) & \mathcal{F}_{\downarrow\downarrow}(\omega)
\end{pmatrix}.
\label{Fmat}
\eea
The diagonal parts of $\mathcal{F}(\omega)$ in Eq.({\ref{Fmat}}) give the information about the equal-spin triplet $s$-wave pair amplitude, while, the off-diagonal components provide the mixed-spin triplet $s$-wave pair amplitude ($\mathcal{F}_{\uparrow \downarrow}+\mathcal{F}_{\downarrow \uparrow}$) and the spin-singlet pair amplitude ($\mathcal{F}_{\uparrow \downarrow}-\mathcal{F}_{\downarrow \uparrow}$). Here each component of $\mathcal{F}(\omega)$ is a  $N\times N$-dimensional matrix, where we extract the pair amplitude in each layer and denote it by $\mathcal{F_{\sigma,\sigma^{\prime}}}(\omega,n)$ for the $n_{\text{FM}}$-th layer of the FM. Finally, we also take the summation over $\omega$ within the SC gap, $|\mathcal{F}_{\sigma,\sigma^{\prime}}(n)|=|\sum\limits_{\omega}\mathcal{F_{\sigma,\sigma^{\prime}}}(\omega,n)|$, to concentrate on the SC gap energy regime in our study. This is done since our main purpose is to isolate the contributions from the proximity-induced pairing. We here use the retarded/advanced Green's functions to be able to capture the even- and odd-frequency dependence of the superconducting pair amplitude, but with the low temperatures found in conventional superconductors, we do not anticipate any significant temperature effects on the pair amplitudes.


\subsection{Results and discussion}
In order to analyze the proximity-induced pair amplitude in the FM, we plot the magnitude of $\mathcal{F}$ as a function of the number of layers of the FM $n_\text{FM}$ for various combinations of $P$, $\Lambda_{\text{RSOI}}$ and $\theta_{\text{F}}$ as presented in Fig.\,\ref{figPA}. We here present the behavior of the pair amplitude for some selected parameter regimes in order to understand the subsequent results on the thermoelectricity. All the possible four pairing: spin-singlet ($\uparrow \downarrow-\downarrow \uparrow$), equal-spin (both $\uparrow\uparrow$ and $\downarrow\downarrow$) triplets and mixed-spin ($\uparrow \downarrow+\downarrow \uparrow$) triplet are considered.

In Fig.\,\ref{figPA}(a), we set $P=0$ and $\Lambda_{\text{RSOI}}=0$ which represents the situation of a normal metal/SC junction. There is only spin-singlet $s$-wave pairing proximity-induced in the FM, as expected when the SC is a conventional $s$-wave spin-singlet SC. The pair amplitude is maximum at the interface and decays almost exponentially as we move towards the inside of the FM. Note that, following $\mathcal{SPOT}=-1$ classification\,\cite{Berezinskii,linderRMP}, any spin-triplet pair amplitude has to be odd-$\omega$ in nature as it is $s$-wave, whereas the spin-singlet pair amplitude is has an even-$\omega$ behavior. In order to distinguish these easily, we introduce the total odd-$\omega$ spin-triplet pair amplitude, defined as $|\mathcal{F}_{\text{odd}}|=|\mathcal{F}_{\uparrow\uparrow}|+|\mathcal{F}_{\downarrow\downarrow}|+|\mathcal{F}_{\uparrow\downarrow+\downarrow\uparrow}|$, where the notation $\mathcal{F}_{\sigma \sigma^{\prime}+\sigma^{\prime}\sigma}$ denotes $\mathcal{F}_{\sigma\sigma^{\prime}}+\mathcal{F}_{\sigma^{\prime}\sigma}$, and plot it with a dotted black line in order to directly compare with the spin-singlet even-$\omega$ pair amplitude, $|\mathcal{F}_{\uparrow \downarrow-\downarrow \uparrow}|$, marked by the dashed red line. 

With the onset of finite polarization and RSOI, spin-triplet pair amplitude appears in the FM region, in Fig.\,\ref{figPA}(b) plotted for the values $P=1$, with $\theta_F=0$ and  $\phi_{\text F}=0$, and $\Lambda_{\text{RSOI}}=0.3$. Since the $P\approx 1$ has predominantly only one spin species present at the Fermi level, we denote this as the half-metallic regime. From Fig.\,\ref{figPA}(b) we see that similar to Fig.\,\ref{figPA}(a), the spin-singlet even-$\omega$ pair amplitude decays gradually with distance from the interface. We check (not shown) that increasing the polarization of the FM results in a gradual reduction in the overall spin-singlet amplitude. We only show the results for $P=1$ to concentrate on the regime where spin-singlet even-$\omega$ is at its minimum. When the polarization of the FM is high, a strong preferential direction along the $z$-axis is set up within the FM region with an asymmetry in the spin DOS and thus, reduces the spin-singlet pair amplitude. The decaying nature of the pair amplitude is accompanied by oscillations inside the ferromagnet, as also reported earlier\,\cite{buzdinRmp}.These oscillations result from the center-of-mass momentum of the Cooper pair acquired due to the ferromagnetic exchange field\,\cite{buzdinRmp}. In addition to the spin-singlet pair amplitude, the odd-$\omega$ pair amplitude is finite in the half-metallic regime. We see that both the equal-spin triplet ($\uparrow\uparrow$ and $\downarrow\downarrow$) pair amplitudes are zero. The only existing triplet pairing in this situation is mixed-spin-triplet ($\uparrow\downarrow+\downarrow\uparrow$) pairing, resulting in a complete overlap of the total odd-$\omega$ pair amplitude to the mixed-spin triplet odd-$\omega$ pair amplitude. The mixed-spin triplet odd-$\omega$ pairing appears because of the spin-mixing induced by the ferromagnetic exchange field\,\cite{JacobPRB,linder2015superconducting,vezin2019enhanced}, whereas, the equal-spin triplets are missing due to the particular choice of the direction of the FM polarization vector, $\theta_F =0$, compared to the direction of the RSOI field\,\cite{halterman}. 
Overall, the total odd-$\omega$ pair amplitude becomes comparable to that of the spin-singlet pairing with a small spatial shift. However, the odd-$\omega$ pair amplitude depends on the magnitude of the polarization and increases with an increase in the polarization of the FM, opposite to the behavior of the even-$\omega$ pairing. 

In the literature, it has previously been shown that the presence of the RSOI at the interface can rotate the quantization axis of the pairing and also generate the equal-spin triplets\,\cite{linderRMP}. This can also be realized in our FM/SC structure by rotating the FM vector with respect to the direction of the fixed RSOI field. We therefore set $\theta_{\text{F}}=\pi/4$, keeping the other parameters the same as in Fig.\,\ref{figPA}(b) and plot the resulting $|\mathcal{F}|$ in Fig.\,\ref{figPA}(c). Similar to the $\theta_{\text{F}}=0$ case, both the spin-singlet and the mixed-spin triplet pair amplitudes are finite within the FM region. Additionally, both equal-spin triplet amplitudes are also present in the FM, with the amplitudes of the $\uparrow\uparrow$ and $\downarrow\downarrow$ equal-spin triplet pairs being equal to each other. As a consequence, the total amplitude of the odd-$\omega$ spin-triplet pair dominates over the even-$\omega$ spin-singlet pair amplitude in most layers of the FM, although there are some minor oscillations. This proximity-induced superconductivity results from the Andreev reflection, which occurs when an electron of a particular spin with an energy within SC gap is incident on the interface and a hole of opposite spin is reflected back. With the increase of the polarization of the FM, the asymmetry in the DOS corresponding to the two different spin bands of the FM gradually increases. This results in the reduction of the Andreev reflection at the interface, but, due to the presence of RSOI, the incident electron may also flip its spin at the interface and take an electron of the same spin and pair together to pass through the SC. This is the phenomenon of spin-flip Andreev reflection, associated with the injection of the equal-spin triplet odd-$\omega$ Cooper pairs in the FM.  
We can also rotate the polarization of the FM further compared to the interface RSOI by setting the direction of $\mathbf{m}$ along $x$-axis, i.e.~$\theta_{\text{F}}=\pi/2$. Then we find zero mixed-spin triplet pair amplitude, whereas, $\uparrow\uparrow$ and $\downarrow\downarrow$ spin-triplet pair amplitude are stronger, as shown in Fig.\,\ref{figPA}(d). Here, the even-$\omega$ and the odd-$\omega$ pair amplitudes become comparable to each other. 

To summarize, there is only proximity-induced even-$\omega$ spin-singlet pairing in the case of no spin polarization in the FM. We have finite mixed-spin triplet odd-$\omega$ pair amplitude in the FM for $P=1$ in the presence of finite RSOI when the polarization vector is parallel to the $z$-axis. Rotation of the polarization vector from the $z$-axis towards the $x$-axis affects both the spin configuration and the amplitude of the odd-$\omega$ pairing proximity-induced in the FM, but the odd-$\omega$ pairs always dominates or are comparable to the spin-singlet pairing. However, the behaviors of all the pair amplitudes are insensitive to the rotation of the $\bm{m}$ vector within the $x-y$ plane (change in $\phi_{\text F}$), since the RSOI field lies on the same plane. 
Note that we here only show the behavior of the $s$-wave pair amplitudes because of its stability against disorder\,\cite{Michaeli}. We check the amplitudes of all possible $p$-wave pairing between the nearest neighbor sites along the three different directions, but they are one or two orders of magnitude smaller than that of the $s$-wave pair amplitude in the half-metal regime in the presence of the RSOI, particularly the regime we are interested in. We thus only discuss the contributions of the proximity-induced $s$-wave pair amplitude to the thermoelectricity throughout this work.

\section{Thermoelectric properties}\label{thermo}
With the understanding of the proximity-induced pair amplitude in the FM, we calculate the themoelectric coefficients for the FM/SC junction for different polarizations both in the absence and presence of RSOI. First we describe the necessary theory followed by our results.

\subsection{Theoretical background} 
In this subsection, we define the thermoelectric coefficients and illustrate the method used to calculate these coefficients numerically.

\subsubsection{Thermoelectric coefficients}
In the linear response regime, the charge current $I_{\text{c}}$ and the thermal current $I_{\text{q}}$ can be expressed following the Onsager matrix equation\,\cite{onsager1931reciprocal1,onsager1931reciprocal2}
\begin{subequations}\label{onsager}
\begin{align}
\label{onsager:a}
I_{\text{c}}&=\mathcal{L}_0 \nabla V + \mathcal{L}_1 \nabla T/T, \\
\label{onsager:b}
I_{\text{q}}&=\mathcal{L}_1 \nabla V + \mathcal{L}_2 \nabla T/T,
\end{align}
\end{subequations}
where $\nabla V$ and $\nabla T$ represent the bias voltage and temperature gradient, respectively.  All the thermoelctric coefficients of Eq.\,\ref{onsager} can be expressed by a general expression,
\bea
\mathcal{L}_{\alpha}=\int\limits_0^{\infty}\int\limits_{S}\, \mathcal{T}(E,\bm{k}_{||}) (E-\mu_{\text{FM}})^{\alpha}\left(-\frac{\partial f}{\partial E} \right)\frac{d^2\bm{k}_{||}}{2 \pi k_F^2} dE,~~~~\label{genL}
\eea
where $\alpha$ is an integer number, being $0$, $1$ or $2$ and indicating $\mathcal{L}_0$, $\mathcal{L}_1$, $\mathcal{L}_2$, known as the electrical conductance, thermoelectric coefficient and thermal conductance, respectively. The transmission function $\mathcal{T}(E,\bm{k}_{||})$ provides the probability of transmission and it is weighted by the energy measured with respect to the Fermi energy \ie $(E-\mu_{\text{FM}})$ in Eq\,.(\ref{genL}). We provide the detailed calculation of $\mathcal{T}(E,\bm{k}_{||})$ later in this subsection. Moreover, $E$ is the energy and $k_F$ is the Fermi wave vector, while the Fermi distribution function is given by $f(E)=1/\left(e^{(E-\mu_{\text{FM}})/k_B T}+1\right)$ with the Boltzmann constant $k_B$. We also perform a surface integration, with area denoted by $S$, in the plane parallel to the interface. 

We can describe other thermoelectric coefficients in terms of the $\mathcal{L}_{\alpha}$ in Eq.~\eqref{onsager} as follows.
The Seebeck coefficient or the thermopower, defined as the open circuit voltage per unit temperature gradient, is found as\,\cite{wysokinski2012thermoelectric,goldsmid2017physics}
\beq
\mathcal{S}=-\mathcal{L}_1/\mathcal{L}_0.
\label{sbk}
\eeq
To calculate the efficiency of the system, we compute the thermoelectric figure of merit $zT$\,\cite{altenkirch1911elektrothermische,goldsmid2017physics}:
\beq
zT=\frac{\mathcal{L}_0 \mathcal{S} T}{\mathcal{L}_2-\mathcal{L}_p}.
\label{zt} 
\eeq
Here, the correction factor due to the Peltier effect is included via the term $\mathcal{L}_p$ and given by\,\cite{bardas1995peltier} 
\beq
\mathcal{L}_p=\frac{\mathcal{L}_1^2}{T \mathcal{L}_0}.
\eeq
All these thermoelectric coefficients can thus be found by calculating the transmission function $\mathcal{T}(E,\bm{k}_{||})$, which we describe next. Note that we have neglected phonon contributions to $\mathcal{L}_2$ by assuming a quasi-equilibrium condition of the phonon distribution function at low temperature, which is typical in $\sim mK$ in experiments\,\cite{kolendaprl2016,kolendaprb2017,giazottormp2006}.

\subsubsection{Transmission function}
We employ the scattering matrix formalism to calculate the transmission function in terms of the probabilities of all the scattering processes occurring at the interface. For this we most easily use the BdG Hamiltonian in Eq.\,(\ref{Hmat}).
We start with the solution of the BdG equations expressed generally as
\bea
\Psi_{\sigma}(\mathbf{r})=\Psi_{\sigma}(z) e^{i (\bm{k}_{||} \cdot\bm{r})}
\eea
following the translational symmetry along $x$- and $y$-directions, with $\bm{k}_{||}$ being that wave vector and the position vector $\bm{r}$ being in the plane of the interface ($x-y$ plane). For the FM ($z>0$) region\,\cite{hogl}, 
\bea
\Psi_{\sigma}^{\text{FM}}(z)&=&\frac{1}{\sqrt{k_{\sigma}^e}} e^{i k_{\sigma}^e z}\psi_{\sigma}^e + r_{\sigma,\sigma}^e
e^{-ik^e_{\sigma}z}\psi_{\sigma}^e+r_{\sigma,\sigma^{\prime}}^e e^{-ik^e_{\sigma^{\prime}}z}\psi_{\sigma^{\prime}}^e \non \\
&&+r_{\sigma,\sigma}^h e^{ik^h_{\sigma}z}\psi_{\sigma}^h 
+r_{\sigma,\sigma^{\prime}}^h e^{ik^h_{\sigma^{\prime}}z}\psi_{\sigma^{\prime}}^h
\label{psi_FM}
\eea
with $k^{e(h)}_{\sigma}=\sqrt{k_F^2-k_{||}^2+(\sigma h_{FM} \pm  2E)/\hbar^2}$ being the electron (hole)-like wave vector. Here, $E$ is the energy of the incoming particle/hole and $\sigma$ is $\pm 1$ if the spin is parallel/anti-parallel to the vector $\mathbf{m}$. Moreover, $r^{e(h)}_{\sigma,\sigma^{\prime}}$ represents the ordinary (Andreev) reflection amplitude with the first and second subscript indicating the spins of the incident and reflected particles, respectively. 
The spinors for the electron-like and hole-like quasiparticles are expressed as
\bea
\psi_{\sigma}^e=\frac{1}{\sqrt{2}}\left[\begin{array}{c}
\sigma\sqrt{1+\sigma\cos{\theta_F}}e^{-i \phi_F} \\ \sqrt{1-\sigma\cos{\theta_F}} \\0 \\ 0\end{array}\right]
\eea
and 
\bea
\psi_{\sigma}^h=\frac{1}{\sqrt{2}}\left[\begin{array}{c}
0 \\ 0 \\ \sigma\sqrt{1+\sigma\cos{\theta_F}}e^{-i \phi_F} \\ \sqrt{1-\sigma\cos{\theta_F}}\end{array}\right],
\eea
which are obtained by diagonalizing Eq.\,(\ref{ham_fm})\,\cite{PDFM}.
Turning to the SC region ($z<0$), the solution is instead given by\,\cite{hogl} 
\bea
\Psi_{\sigma}^{\text{SC}}=t^e_{\sigma,\sigma}\left[\begin{array}{c} u
\\ 0 \\v\\0\end{array} \right]{e}^{iq_{e}z}+ t^e_{\sigma,\sigma^{\prime}}\left[\begin{array}{c} 0
\\ u \\0\\v\end{array} \right]{e}^{iq_{e}z}\non \\
+t^h_{\sigma,\sigma}\left[\begin{array}{c} u
\\ 0 \\v\\0\end{array} \right]{e}^{-iq_{h}z}+ t^h_{\sigma,\sigma^{\prime}}\left[\begin{array}{c} 0
\\ u \\0\\v\end{array} \right]{e}^{-iq_{h}z},
\label{psi_SC}
\eea
where $q_{e(h)}=\sqrt{q_F^2-k_{||}^2 \pm 2 \sqrt{E^2-\Delta_{\text{SC}}^2}/\hbar^2}$ with the SC coherence factors given by $u(v)=\sqrt{[1\pm\sqrt{1-\Delta^2_{\text{SC}}/E^{2}}]/2}$. Here, $t_{\sigma,\sigma^{\prime}}^{e(h)}$ denotes the amplitude of transmitted electron (hole)-like quasiparticles and $q_F$ denotes the Fermi wave vector within SC region. Similar to the reflection coefficients, the first and second subscript denote the spins of the incident and the transmitted particles, respectively. We note that in this formalism all superconducting and temperature effects only enter through the superconducting order parameter $\Delta_{\text{SC}}$. In particular, we do not have to explicitly enter into the calculation any of the calculated pair amplitudes in the FM region in the previous section, but they are indirectly incorporated.

Next, we employ the Andreev approximation\,\cite{andreev} to neglect all higher order energy terms since we are interested in the low energy contributions only. Thus we can express the wave vectors in the FM region as $k^e_{\sigma}\approx k^h_{\sigma}\approx k_F \sqrt{1+\sigma P -k^2}$ and in the SC region as $q^e\approx q^h\approx k_F\sqrt{(q_F/k_F)^2-k^2}$ where $k$ is a dimensionless wave vector defined as $k_{||}/k_F$.
We finally use two boundary conditions, which maintains the continuity of both the wave function and the momentum at the interface:
\bea
\Psi^{\text{FM}}_{\sigma}|_{z=0}=\Psi_{\sigma}^{\text{SC}} |_{z=0}&&\non\\
\frac{\hbar^2}{2}\left(\frac{d}{dz}\Psi_{\sigma}^{\text{SC}} |_{z=0}-\frac{d}{dz}\zeta\Psi_{\sigma}^{\text{FM}}|_{z=0}\right)  \non \\
=\sigma_3 \otimes (\mathbf{W}_{\text{RSOI}}.\bm{\sigma})~\Psi_{\sigma}^{\text{FM}}|_{z=0}
\eea
 where $\zeta=diag(1,1,-1,-1)$, to find the scattering coefficients of Eqs.\,(\ref{psi_FM}) and (\ref{psi_SC}). Note that the effect of the interfacial RSOI enters through the boundary condition\,\cite{PDFM}.   
 In the end, the transmission function $\mathcal{T}(E,k_{||})$ can be expressed in terms of the ordinary and Andreev reflection probability using Blonder-Tinkham-Klapwijk formalism\,\cite{blonder}
\beq
\mathcal{T}(E,k_{||})=\sum\limits_{\sigma}[1-R^e_{\sigma}(E,k_{||})+R^h_{\sigma}(E,k_{||})],
\label{teqn}
\eeq
where the ordinary and Andreev reflection probabilities can be found from the relation $R_{\sigma}^{e(h)}(E,k_{||})=\text{Re}[k_{\sigma}^{e(h)}|r_{\sigma}^{e(h)}|^2+k_{-\sigma}^{e(h)}|r_{-\sigma}^{e(h)}|^2]$, following the current conservation at the junction. 
 
\subsection{Results and discussion}
Having developed the necessary theoretical framework, we present our results of the Seebeck coefficient and thermoelectric figure of merit for FM/SC junction and also discuss the role of the proximity-induced pair amplitude. There are both even-$\omega$ and odd-$\omega$ pair amplitudes present in the system. To as much as possible isolate the contribution of the odd-$\omega$ pairing, we only focus on the regimes for which the ratio of odd- to even-$\omega$ pair amplitude is high and relate the results of the thermoelectric coefficients to the results of the pair amplitude discussed in the previous section.

\begin{figure}[htb]
\includegraphics[scale=0.57]{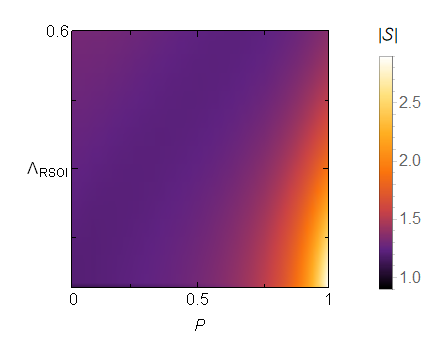}
\caption{Color plot of the magnitude of Seebeck coefficient $|\mathcal{S}|$ in units of $k_B/e$ as a function of the polarization $P$ of the FM and the interfacial RSOI strength $\Lambda_{\text{RSOI}}$, keeping $\theta_{\text F}=0$.}
\label{seebeck3d}
\end{figure}

In Fig.\,\ref{seebeck3d}, we show a color plot of the Seebeck coefficient amplitude $|\mathcal{S}|$ as a function of the polarization of the FM and RSOI strength $\Lambda_{\text{RSOI}}$ by setting $\theta_{\text{F}}=0$ and $\phi_{\text{F}}=0$. The dark to light color represents the low to high magnitude of $|\mathcal{S}|$. We here ignore the 
sign of $\mathcal{S}$ as it is always negative for all the parameter values in our study. We see that for $\Lambda_{\text{RSOI}}=0$ and $P=0$, where we have only the conventional $s$-wave spin-singlet even-$\omega$ pairing proximity-induced in the FM, see Fig.\,\ref{figPA}(a), the value of $|\mathcal{S}|$ is low. With the increase of the polarization of the FM, Seebeck coefficient increases. Particularly, in the half-metal regime ($P\approx 1$), where strong spin-triplet odd-$\omega$ pairing is proximity-induced along with the spin-singlet even-$\omega$ pairing, we get a noticably higher $|\mathcal{S}|$ in the junction. This is true both in the absence and presence of finite low to moderate RSOI. 

We note that in the numerator of the Seebeck coefficient, see Eq.\,(\ref{genL}) and (\ref{sbk}), the term $(E-\mu_{\text{F}M}) (\partial f/\partial E)$ is an odd function of $E$. This means that only the odd in $E$ part of the transmission function contributes to $\mathcal{L}_1$ after integrating over $E$\,\cite{sothman}, whereas, both the even and odd in $E$ part of $\mathcal{T}(E)$ contribute to $\mathcal{L}_0$. Consequently, only the odd-$\omega$ part of the anomalous Green's function, or equivalently pair amplitude, contributes to $\mathcal{L}_1$. The even in $E$ part of $\mathcal{T}(E)$, or  equivalently even-$\omega$ pair amplitude, thus always reduces the Seebeck coefficient, since there is no sign change in between the contributions by the even-$\omega$ and odd-$\omega$ pair amplitude. All the above arguments leads to the conclusion that odd-$\omega$ pairing is more effective in enhancing the Seebeck voltage. To further reinforce our prediction, we analyze our numerical results.
In particular, we relate the results of the thermoelectric coefficients to the pair amplitudes for the various parameter regimes presented in Figs.\,\ref{figPA}(b)-\ref{figPA}(d). In Fig.\,\ref{seebeck3d} in the $P\approx1$ regime, the Seebeck coefficient in highest for $\Lambda_{\text{RSOI}}=0$ but also high in presence of low RSOI. It is here necessary to not focus on $\Lambda_{\text{RSOI}}=0$ where $\mathcal{S}$ is highest. Instead, we concentrate on the case of low RSOI, keeping $P=1$, the reason is as follows.

When $\Lambda_{\text{RSOI}}=0.3$ and $\theta_{\text F}=0$, $\mathcal{S}$ is high and we have both even-$\omega$ spin-singlet and odd-$\omega$ spin-triplet pairing, see Fig.\,\ref{figPA}(b), in the system. There may be contributions by both the subgap and the supergap energy levels to the Seebeck coefficient. Turning our attention to the contribution by the proximity-induced pairing, or equivalently supercurrent, we need to specifically focus on the SC subgap energy regime by setting the limit of the integration over the energy in Eq.\,(\ref{genL}) to $\Delta_{\text{SC}}$, as this avoid contributions from the quasiparticles present above the gap. 

\begin{figure}[htb]
\includegraphics[scale=0.5]{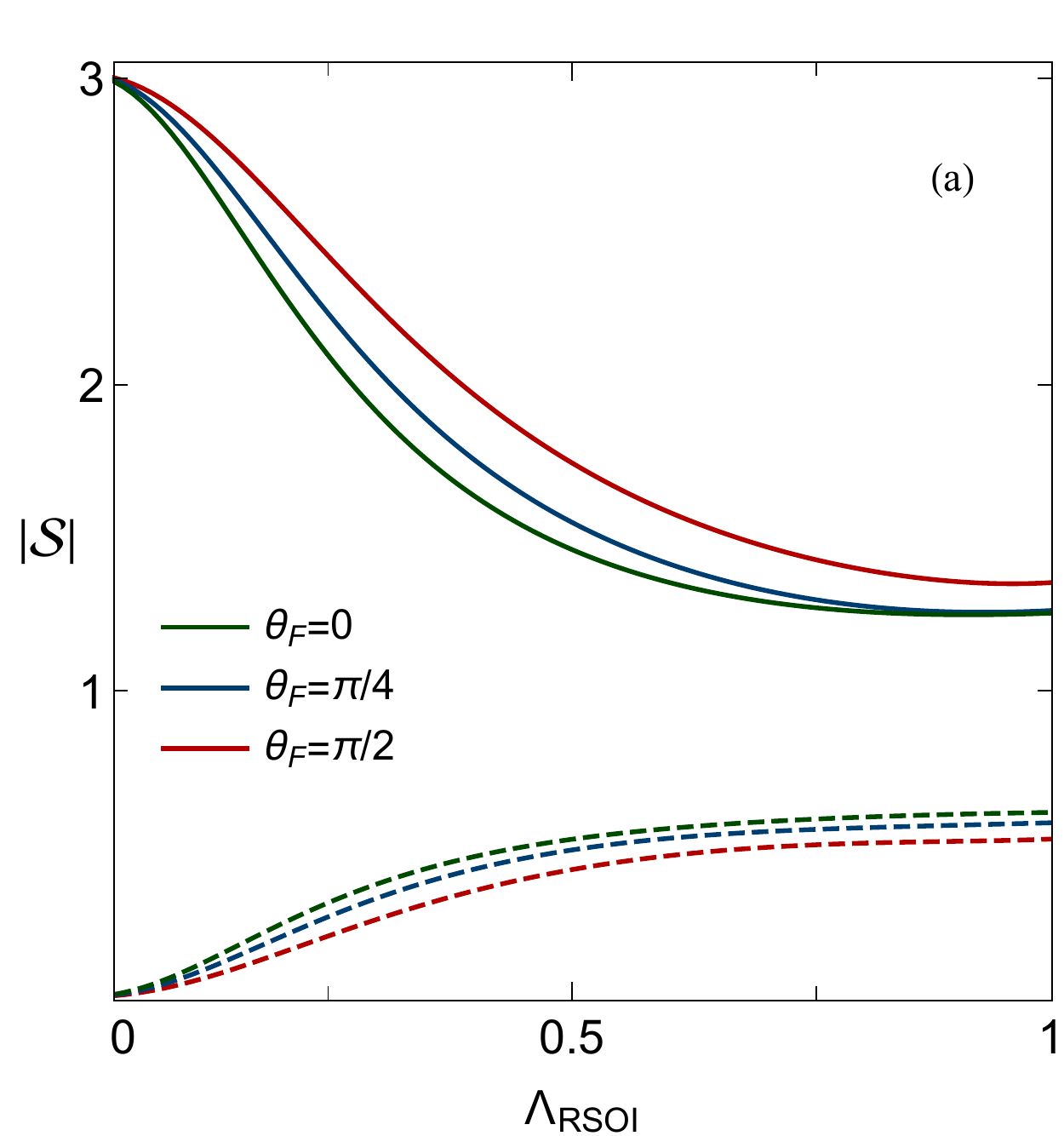}
\includegraphics[scale=0.5]{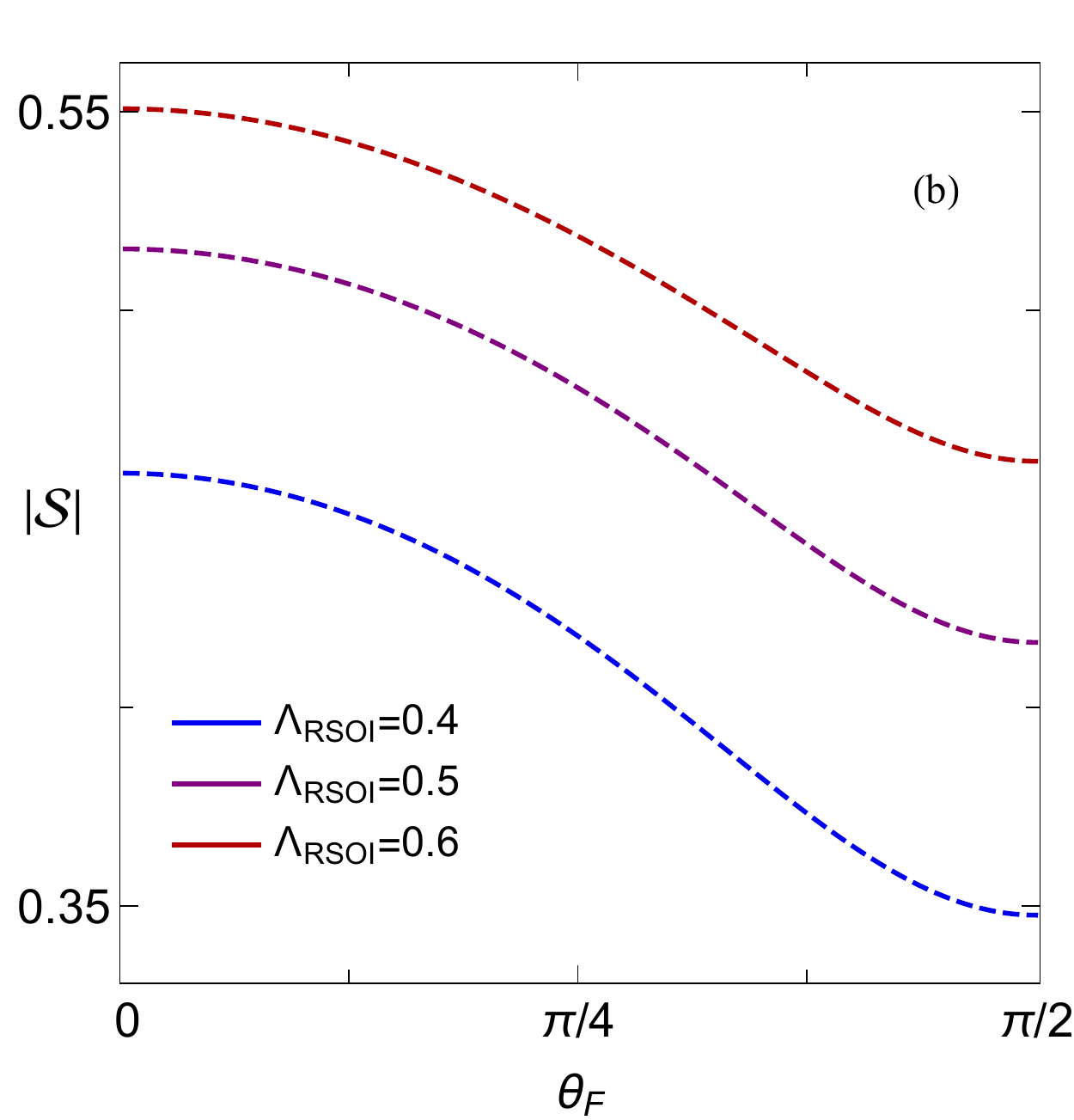}
\caption{Magnitude of the Seebeck coefficient $|\mathcal{S}|$ in units of $k_B/e$ for $P=1$ as a function of (a) $\Lambda_{\text{RSOI}}$ for all the energy levels (solid) and considering only the subgap energy levels (dashed) and (b) the polar angle $\theta_{\text F}$ of the polarization vector $\bm{m}$ for $\phi_{\text{F}}=0$ and RSOI strength $\Lambda_{\text{RSOI}}=0.3$ considering only subgap energies. }
\label{Stheta}
\end{figure}

In Fig.\,\ref{Stheta}(a) we plot $|\mathcal{S}|$ as a function of $\Lambda_{\text{RSOI}}$ for $P=1$ of the FM/SC junction, with both the whole range of integration and only SC subgap energy limit for three different $\theta_{\text{F}}$ values.
For $\theta_{\text{F}}=0$, as in Fig.\,\ref{seebeck3d}, we observe that the magnitude of the Seebeck coefficient is highest for $\Lambda_{\text{RSOI}}=0$ when we consider all the energy levels including the subgap regime. In contrast, when we plot $|\mathcal{S}|$ by taking only the contributions from the subgap energies, we see that the subgap contribution is zero at this field orientation. Thus for $\Lambda_{\text{RSOI}}=0$, the only contribution is from the energy levels above the SC gap. It is thus the tunneling processes occurring at the interface, that are responsible for the thermoelectricity. Hence, the transmitted electron-like and hole-like quasiparticles are the only carriers of the thermal current in this parameter regime. Specifically, there is no contribution by the proximity-induced pair amplitude for $\Lambda_{\text{RSOI}}=0$. This is true for any finite $\theta_{\text{F}}$ values,  as shown for $\theta_{\text F}=\pi/4$ and $\pi/2$. Thus, we can here disregard the $\Lambda_{\text{RSOI}}=0$ case from our discussion as our focus is the contribution by the proximity-induced odd-$\omega$ pair amplitude. 

In Fig.\,\ref{Stheta}(a) the total Seebeck coefficient $|\mathcal{S}|$ is reduced and the subgap contribution forms a significant portion of the total Seebeck coefficient at larger RSOI. Focusing on this subgap regime, we see that the subgap contribution to the Seebeck coefficient significantly increases with the increase of RSOI, although it saturates after certain value of $\Lambda_{\text{RSOI}}$. This shows that the contributions by the proximity-induced pairing increases for low to moderate RSOI. For the situation presented in Fig.\,\ref{figPA}(b), there could, in principle, have been finite contributions by both even-$\omega$ and odd-$\omega$ pairing to the Seebeck coefficient, but we already know that even-$\omega$ spin-singlet pairing is  not good carrier for the thermoelectric current from the result in $P=0$ junctions. Thus we infer that there is a major contribution to the Seebeck coefficient from the odd-$\omega$ pair amplitude, taking place for finite RSOI and at subgap energies.

The only remaining issue with the conclusion that odd-$\omega$ pairing is responsible for the subgap contributions is if there are also subgap states in the junction. In the literature it has previously been shown that within the SC gap, there can be zero energy states due to odd-$\omega$ superconductivity\,\cite{asano2008,foglstrom,di2015a}. We have checked that this is also true in our case in the presence of the finite RSOI for the half-metallic regime of the FM. So, the subgap contribution is generated not only by the supercurrent but also by subgap quasiparticles. But, since the appearance of these zero-energy states are a direct consequence of the odd-$\omega$ pairing, the conclusion still holds that odd-$\omega$ pairing gives a significant contribution to the Seebeck coefficient. 
  
To extract the contributions from the proximity-induced odd-$\omega$ spin-triplets with different spin-structures, we plot in Fig.\,\ref{Stheta}(b) $|\mathcal{S}|$ as a function of $\theta_{\text F}$, keeping $\phi_{\text F}=0$. With the increase of $\theta_{\text F}$, the magnitude of $\mathcal{S}$ decreases. Thus, the case where the polarization vector $\bm{m}$ lies along the $z$-axis ($\theta_{\text F}=0$), where we have only mixed-spin triplet pairing present in the FM, the magnitude of the Seebeck coefficient is larger compared to that for finite $\theta_{\text F}$. With increasing $\theta_{\text F}$, the equal-spin triplet pair amplitudes are growing but it results instead in somewhat of a reduction in $\mathcal{S}$. 
In particular, when we rotate the polarization vector to $\theta_{\text F}=\pi/4$, the Seebeck coefficient decreases slowly although the total odd-$\omega$ pair amplitude increases by small amount. Thus the decrease in $|\mathcal{S}|$ must be related to the change in the spin-structure of the dominating pair amplitude, i.e.~the reduction in the mixed-spin triplet pair amplitude and the increase of equal-spin triplet pairing. When we rotate the polarization vector further and set $\theta_{\text F}=\pi/2$, the decreasing nature of the Seebeck coefficient still continues and it corresponds to only equal-spin triplet pair amplitude with the mixed-spin triplet pair amplitude being zero. These trends are present for any low to moderate RSOI  (note the y-axis scale). We also check that the change in the angle ($\theta_{\text F}=\pi/4$ to $\pi/2$) does not affect the zero energy states. The mixed-spin-triplet pairing must therefore be more efficient than the equal-spin triplet pair amplitude in generating a thermoelectric voltage at the FM/SC junction.

\begin{figure}[htb]
\includegraphics[scale=0.5]{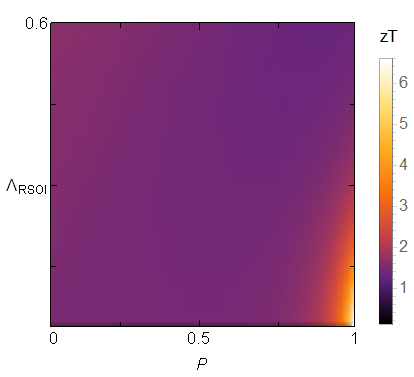}
\caption{Thermoelectric figure of merit ($zT$) as a function of the polarization $P$ and RSOI for the polar angle $\theta_{\text F}=0$ of the polarization vector $\bm{m}$ of the FM.}
\label{figzT3d}
\end{figure}
Finally, we also calculate the thermoelectric figure of merit and plot it as a function of polarization of the FM and RSOI in Fig.\,\ref{figzT3d} for $\theta_{\text F}=0$. We see that, similar to the Seebeck coefficient, there is an enhancement in $zT$ in the half-metal regime where odd-$\omega$ pairing is proximity-induced in the FM. For the half metal FM, in the presence of low RSOI, $zT$ even increases to a value of $5$. Notably this is much higher than in the regime of zero RSOI and zero polarization, i.e.~the normal metal regime where only spin-singlet even-$\omega$ paring is present and the thermoelectric effect is only due to transmitted quasiparticles. The FM/SC junction is thus a much more efficient thermoelectric junction when the proximity-induced odd- to even-$\omega$ pair amplitude is high in the system compared to the situation when only even-$\omega$ pairing is induced in the FM. 

The thermoelectric figure of merit $zT$ reaches maximum value when the FM is in the half-metal regime $P\approx1$, where mostly only one of the spin spices dominates the density of states at Fermi level. Either the absence of RSOI or the presence of low RSOI is the favored condition to achieve the highest $zT$. Similar to the Seebeck coefficient, we concentrate on the regime where we have high odd-$\omega$ pair amplitude and thus avoid the zero RSOI case. Physically, the combination of half-metallic regime and low RSOI instigate spin-flip Andreev reflections at the interface. Spin-flip Andreev reflection occurs when an electron of a particular spin with an energy within SC gap is incident on the interface, flip its spin due to the broken spin-symmetry (happens because of RSOI in this case) at the junction, and a hole of same spin (as the incident electron) is reflected back. It enhances the probability of inducing odd-$\omega$ pairing, and thus more odd-$\omega$ pairing is proximity-induced in the system. Enhancement of $zT$ strongly supports our prediction from the behavior of Seebeck coefficient that odd-$\omega$ pairs are good carriers of thermoelectricity in SC/FM junction. We do not further show the behavior of $zT$ with the rotation of the FM polarization vector as it is very similar to the behavior of $\mathcal{S}$. 

\section{Summary} \label{conclu}
We have investigated the role of odd-$\omega$ pairing in the thermoelectricity of a FM/SC junction with an applied temperature gradient. We have done this by calculating the proximity-induced pair amplitude and several thermoelectric coefficients, the Seebeck coefficient and thermoelectric figure of merit, for various conditions of the interface and the polarization of the FM region. We have found that the odd-$\omega$ spin-triplet pairing is playing a major role in the thermoelectricity of the FM/SC junction in the presence of a spin-active interface, in particular the mixed-spin triplet pairing. In fact, the efficiency of the FM/SC junction as a thermoelectric device is highly enhanced when odd-$\omega$ pairing is present compared to the situation when there is only even-$\omega$ pairing in the system. 

For the experimental realization of thermoelectricity in FM/SC junctions, we mention some parameter values as example. Using a BCS SC (for example, Nb with $T_c=9.3K$ and $\Delta_0\sim 1.45$ meV) we can have proximity-induced SC gap $\sim 80\%$ of the original SC gap\,\cite{chrestin}. In presence of a ferromagnetic insulator like europium sulfide, we can get higher thermoelectricity ($\mathcal{S}\sim 100\mu$V/K) by using small exchange field ($\sim$mT)\,\cite{kolendaprb2017}. Keeping the base temperature in sub-Kelvin order, a thin layer of zincblende can be used to get interfacial RSOI $\sim 60$ meV \AA\,\cite{manchon2015}, which corresponds to the high $zT$ regime.

\acknowledgments{We acknowledge J.~Cayao and Andreas Theiler for useful discussions and financial support from the Knut and Alice Wallenberg Foundation through the Wallenberg Academy Fellows program, the Swedish Research Council (Vetenskapsr\aa det Grant No. 2018-03488), the European Research Council (ERC) under the European Unions Horizon 2020 Research and Innovation programme (ERC-2017-StG-757553), and the EU-COST Action CA-16218 Nanocohybri.}


\bibliography{bibfile}{}

\end{document}